\documentclass[10pt,twocolumn,letterpaper]{article}

\usepackage{acpr}
\usepackage{times}
\usepackage{epsfig}
\usepackage{graphicx}
\usepackage{amsmath}
\usepackage{amssymb}

\usepackage[T1]{tipa} 
\usepackage{multirow}
\usepackage{mathtools}
\usepackage{enumerate}
\usepackage{caption}
\usepackage{subfigure,ctable}
\usepackage{bm} 
\usepackage{makecell}

\DeclareMathOperator*{\argmax}{arg\,max}

\def\baselinestretch{0.98}

\newcommand{\myfigureshrinker}{\vspace{-0.5cm}}

\usepackage[pagebackref=true,breaklinks=true,letterpaper=true,colorlinks,bookmarks=false]{hyperref}

\acprfinalcopy 

\def\acprPaperID{499} 

\ifacprfinal\fi

\begin{document}

\title{Accent Classification with Phonetic Vowel Representation}

\author{Zhenhao Ge, Yingyi Tan, Aravind Ganapathiraju\\
Interactive Intelligence Inc., Indianapolis, Indiana, USA\\
{\tt\small \{roger.ge, yingyi.tan, aravind.ganapathiraju\}@inin.com}
}

\maketitle
\thispagestyle{empty}

\begin{abstract}

Previous accent classification research focused mainly on detecting accents with pure acoustic information without recognizing accented speech. This work combines phonetic knowledge such as vowels with acoustic information to build Guassian Mixture Model (GMM) classifier with Perceptual Linear Predictive (PLP) features, optimized by Hetroscedastic Linear Discriminant Analysis (HLDA). With input about 20-second accented speech, this system achieves classification rate of 51\% on a 7-way classification system focusing on the major types of accents in English, which is competitive to the state-of-the-art results in this field.  

\end{abstract}

\vspace{-0.1cm}
\section{Introduction}
\label{sec:introduction}

Improving speech recognition for accented speakers is becoming increasingly more important as businesses become more international. However, handling calls with accents is still a major challenge for companies specializing in speech recognition support services. It requires an accurate and efficient accent classification algorithm, which can identify the accent of the call during a short amount of time, after which, an accent-adapted speech recognition engine can be employed to better recognize accented speech. 

Accent classification recently gains more interests, probably due to the increasing demands for better speaker recognition with accented speech. Recently, Choueiter et al. achieved 32\% classification rate on 23-way classification of accented English \cite{choueiter2008empirical}, using methods such as Maximum Mutual Information (MMI) training and Gaussian tokenization.  Omar et al. used Support Vector Machine (SVM) classifier integrated with Universal Background Model (UBM) and claimed they outperformed the results in \cite{choueiter2008empirical} by 75.3\% relatively \cite{omar2010novel}. Another work in \cite{macias2003acoustic} reported classification rates of 73\% and 58.9\% for German vs. Spanish classification using Gaussian Mixture Models (GMMs) and naive Bayes classification respectively. In addition, classification rates of 36.2\%, 17.7\% and 13.2\% were reported for 4-, 13- and 23-way classification using naive Bayes. To the best of our knowledge, these are the only three works, which used the same dataset as we used in this work.

In this paper, a baseline accent classifier is created using GMMs with purely acoustic features, such as Perceptual Linear Predictive (PLP), which were then discriminatively optimized by Heteroscedastic Linear Discriminant Analysis (HLDA). Based on the fact that most of accents are presented from the pronunciation of vowels rather than consonants, for each type of accents, various GMMs are computed using the same PLP-HLDA features for the vowels extracted from speech. Then, these GMMs are combined to form a single GMM. With the partial transcription of the database for 7 major types of accents, which is absent in \cite{choueiter2008empirical} and \cite{macias2003acoustic}, the vowels are extracted and with these phonetic information, the classification rate of 7-way classification is improved from 46\% to 51\%, compared with the baseline. 

This work was initiated during the first author's internship at Interactive Intelligence \cite{ge2013mispronunciation}. The algorithm and experiment was later refined for better accuracy and efficiency. The following sections are organized as follows: Sec. \ref{sec:data} introduces the database and features used here; the main concept of creating accent-adapted features based on vowel representation is illustrated in Sec. \ref{sec:vowel_representation}; in Sec. \ref{sec:baseline} and \ref{sec:improved}, the implementation and results for the  baseline GMM-HLDA classier and the improved accent classifier with vowel extraction and representation are described in details, followed by the summary and future work in Sec. \ref{sec:conclusion}.

\vspace{-0.1cm}
\section{Data Preparation}
\label{sec:data}

The database used here for developing accent classifiers is Foreign Accented English (FAE) corpus. It was originally collected by the Center of Speech \& Language Understanding (CSLU) at Oregon Health \& Science University (OHSU). It contains 4925 sentences about 20 seconds long each, from speakers with 23 types of accents. 

%
%
We group them into 7 regional accents and one type of accents in each group was selected for developing a 7-way accent classifier, including Arabic (AR), Brazilian Portuguese (BR), French (FR), German (GE), Hindi (HI), Mandarin (MA) and Russian (RU). Tab. \ref{tab:fae_summary} provides a summary of these accents with the number of utterances in each type and their proportion of the entire FAE corpus. In order to perform phoneme alignment which is necessary to extract features with phonetic information, we also transcribed the audio data of these 7 major accents, which is originally absent in the LDC's release. 
\begin{table}[tb]
\centering
\footnotesize
\caption{Summary of selected accents in FAE corpus}
\label{tab:fae_summary}
\renewcommand{\tabcolsep}{1pt}
\begin{tabular}{@{} cccccc @{}} \toprule
Accents & No. of & Proportion & Total & Total & Comp. \\
(Abbr.) & utterances & (\%) & Duration$_1$ & Duration$_2$ & rate (\%) \\ \midrule
AR & 112 & 2.27 & 0:34:32 & 0:29:11 & 84.5 \\
BP & 459 & 9.32 & 2:34:24 & 2:09:58 & 84.2 \\
FR & 284 & 5.77 & 1:31:05 & 1:18:44 & 86.4 \\
GE & 325 & 6.6 & 1:36:04 & 1:22:18 & 85.7 \\
HI & 348 & 7.07 & 1:56:10 & 1:36:31 & 83.1 \\
MA & 282 & 5.73 & 1:30:37 & 1:16:06 & 84.0 \\
RU & 236 & 4.79 & 1:11:13 & 0:59:54 & 84.1 \\
\bottomrule
\multicolumn{6}{p{0.8\linewidth}}{Note: duration$_1$ and duration$_2$ are the duration before and after silence removal}

\end{tabular}
\end{table}


Data from these accents were then preprocessed with silence removal by thresholding on its short-time energy rate and spectral centroids, using method in \cite{giannakopoulos2009method}. Given the audio samples $s_i(n), n \in [1,N]$ in the $i^\textrm{th}$ frame, its short-time energy rate, denoted as $e_i$, can be formulated as
\begin{equation}
e_i = \frac{1}{N}\sum_{n=1}^{N}|s_{i}(n)|^{2} ,
\end{equation}  
where $N$ is the number of samples in one frame. The spectral centroid can be defined as
%
\begin{equation}
c_{i} = \sum_{k=1}^{K}(k+1)S_{i}(k)/\sum_{k=1}^{K}S_{i}(k) ,
\end{equation}
where $S_{i}(k), k \in [1,K]$ is the Discrete Fourier Transform (DFT) coefficients of $s_{i}$. The short-time energy rate is the most useful feature to discriminate silence with environmental noise from speech, and the spectral centroids can be used to remove non-speech noise, such as coughing, due to its lower energy concentration in the spectrum, ralative to that of regular human speech.
\begin{figure}[tb]
  \centering
 	\includegraphics[height=4.5cm, width=7cm]{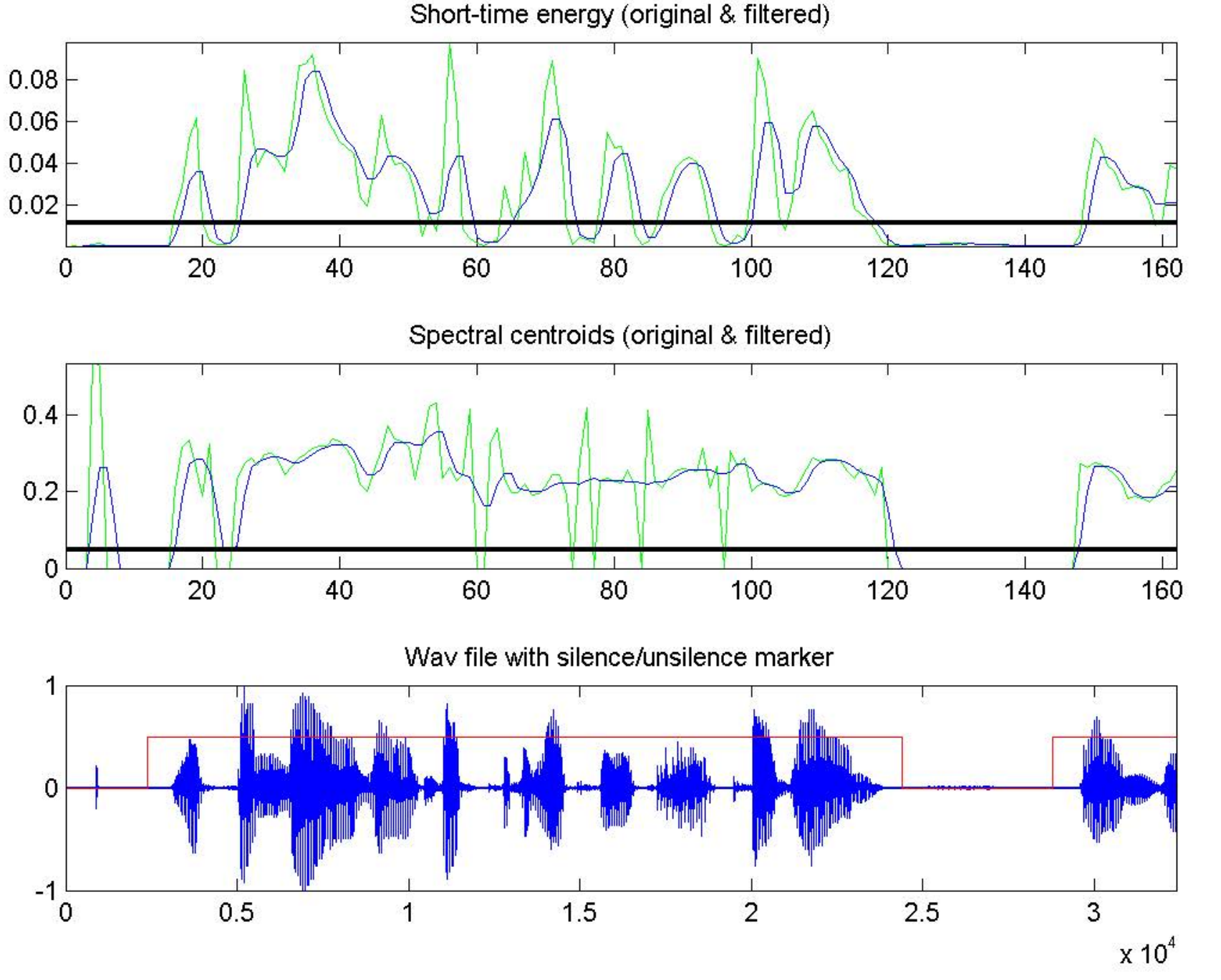}
    \caption{Example of silence removal using short-time energy rate and spectral centroids (FAR00042.wav in FAE) \label{fig:silence_removal_example}}
    \myfigureshrinker
\end{figure} 
Fig. \ref{fig:silence_removal_example} demonstrates the silence removal using both measurements on file FAR00042.wav in FAE corpus with Arabic accents. The portion of speech is considered to be silence when either the smoothed short-time energy rate and the smoothed spectral centroids are below certain thresholds. Tab. \ref{tab:fae_summary} shows the total durations before and after silence removal and their corresponding compression rate.


The silence-removed data of the selected accents were then converted to 39-dimensional PLP features \cite{plp}. Feature Mean and Variance Normalization (MVN) were applied afterwards. They were randomly divided into training, development and testing with ratio $70:15:15$.  

\vspace{-0.1cm} 
\section{Vowel Representation}
\label{sec:vowel_representation}

\begin{figure}[tb]
  \centering
 	\includegraphics[scale=0.28]{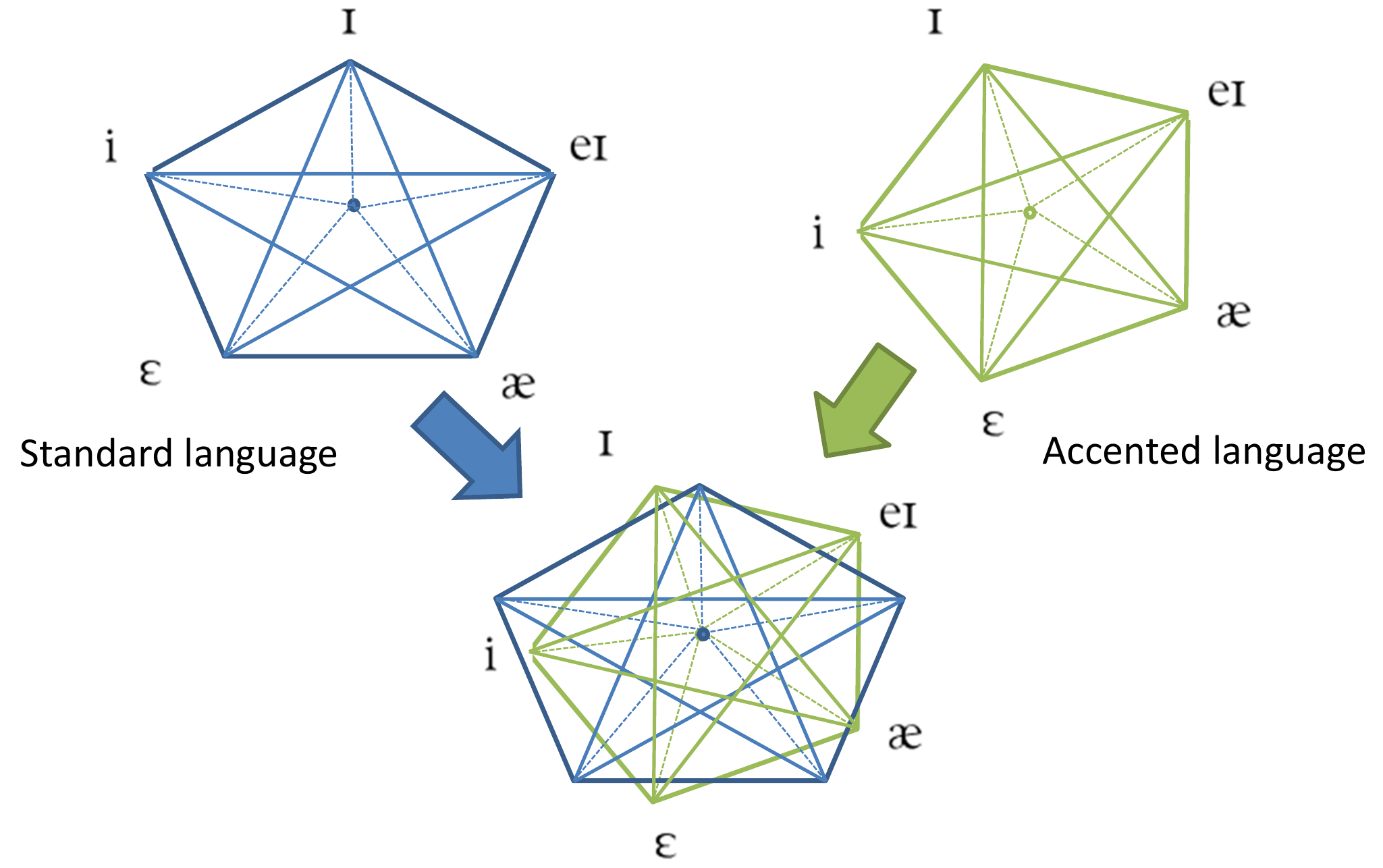}
    \caption{Comparison of 5 vowels locations in standard and accented language\label{fig:5vowels}}
    \myfigureshrinker
\end{figure}

Inspired by the work from Minematsu et al. \cite{Minematsu:2004} and Suzuki et al. \cite{Suzuki:2009}, where they measured the overall structure of the speaker's phonetic space, one type of accent-adapted features can be obtained by extracting vowels from speech and use them to identify accents. For each type of accented version of a target language, such as English, as well as the standard one, it is assumed that the features of the five fundamental vowels are located relatively constantly in the feature space. In Fig. \ref{fig:5vowels}, the first two feature dimensions are taken to illustrate the position of five vowels in accented and non-accented (standard) languages \cite{Minematsu:2004}. The center in each pentagon is the weighted average of five vowels based on their positions in feature space and frequency of appearance in the corpus. By matching the center of the pentagon of the standard and the accented language into the overlapped pentagon in the bottom of Fig. \ref{fig:5vowels}, the Bhattacharyya distances \cite{Bhattacharyya:1946} between each pair of corresponding vowels and their angles can be computed and stored in a vector. This vector $V_i$ represents the difference from the accented language $L_i$ to the standard one $L$. To classify the test speech into one of the accent categories $L_1, L_2, \ldots, L_N$, where $N$ is the number of accents, the difference from $V_j$ to $V_i, i \in [1,N]$ and $V$ (category of the standard language) are computed and classified to the nearest category of accent. 

\vspace{-0.1cm}
\section{Baseline with Pure Acoustic Information}
\label{sec:baseline}

As mentioned in Sec. \ref{sec:introduction}, the baseline accent classification system is implemented using GMM classifier with PLP feature discriminatively optimized by HLDA, which is a generalization of LDA allowing features to have different variances in different feature dimensions. Here we briefly describe the key components of GMM, LDA and HLDA, then discuss the implementation and results of the baseline.

\subsection{GMM Classifer}
\label{subsec:gmm}

Motivated by the method of modeling attributes of speakers using Gaussian Mixture Models (GMMs) in \cite{ge2012pca}, here we use GMMs to model the attributes of accents. Gaussian mixture density models the feature distribution of each accent as a weighted sum of multiple Gaussian distributions. Given row feature vector $\mathbf{x}$ in $M \times K$ feature matrix $X$, where $M$ is feature dimension and $K$ is the number of feature vectors, in the probability of $\mathbf{x}$ can be formulated as 
\begin{equation} \label{eq:gmm_p}
	p(\mathrm{x}|\lambda) = \sum^N_{i=1} w_i b_i(\mathbf{x}), 
\end{equation}
where $N$ is the number of mixture components, and
\begin{equation} \label{eq:gmm_b}
b_i(\mathbf{x}) = \frac{1}{(2\pi)^{M/2}\vert\Sigma_i\vert^{1/2}} \mathrm{exp}\lbrace - \frac{1}{2}(\mathbf{x}-\mathbf{\mu}_i)^T\Sigma^{-1}_i(\mathbf{x}-\mathbf{\mu}_i)\rbrace.
\end{equation}
$b_i(\mathbf{x})$, $i = 1, \ldots, N$, are the component densities, $w_i$ are the mixture weight for $i^{\textrm{th}}$ mixture, and $\lambda = \{w_i,\mu_i,\Sigma_i$\} is the collective representation of the parameters.

Given feature matrix $X$ of accent type $s$, Maximum Likelihood Estimation (MLE) is used to maximize the GMM likelihood, which can be written as 
\begin{equation} \label{eq:mle}
	\lambda^* = \argmax_\lambda p(\mathbf{X}|\lambda) 
	= \argmax_\lambda \prod^K_{k=1}p(\mathbf{x}_k|\lambda).
\end{equation}
Since this expression is non-linear and direct maximization is difficult, the parameter set $\lambda = \{w,\mu,\Sigma\}$ is iteratively estimated using a special case of the Expectation-Maximization (EM) algorithm and is summarized below:  
\begin{eqnarray} \label{eq:gmm}
	\bar{w}_i &=& \frac{1}{K}\sum^{K}_{k=1}p(i|\mathbf{x}_k,\lambda); \nonumber \\
	\bar{\mathbf{\mu}}_i &=& \frac{\sum^{K}_{k=1}p(i|\mathbf{x}_k,\lambda)\mathbf{x}_k}{\sum^{K}_{k=1}p(i|\mathbf{x}_k,\lambda)}; \\
	\bar{\mathbf{\sigma}}^2_i &=& \frac{\sum^{K}_{k=1}p(i|\mathbf{x}_k,\lambda)\mathbf{x}^2_k}{\sum^{K}_{k=1}p(i|\mathbf{x}_k,\lambda)}-\bar{\mathbf{\mu}}^2_i, \nonumber
\end{eqnarray}
where $\bar{w}_i,\bar{\mathbf{\mu}}_i,\bar{\mathbf{\sigma}}^2_i$, $i = 1,...,N$ are the mixture weights, means, and variances for the $i$th component; $p(i|\mathbf{x}_k,\lambda)$ is the \textit{a posteriori} probability for the $i$-th component given by 
\begin{equation} \label{eq:posteriori}
	p(i|\mathbf{x}_k,\lambda) = \frac{w_{i}b_{i}(\mathbf{x}_k)}{\sum^{N}_{j=1}w_{j}b_{j}(\mathbf{x}_k)}\;.
\end{equation}
These estimates are based on the assumption of independence among feature dimension, so for each accent type $s$, the non-zero values of the covariance matrix are only on the diagonals. This algorithm guarantees a monotonic increase of the model's likelihood on each EM iteration. 

After obtaining the GMM parameter set $\lambda_s$ for accent class $s \in [1,S]$, the GMM-based classifier, which maximize \textit{a posteriori} probability for feature matrix $X$ is:
\begin{eqnarray} \label{eq:gmmclassifier}
\small
	\hat{S} & = & \argmax_{s \in [1,S]}p(\lambda_s|X) = \argmax_{s \in [1,S]}\frac{p(X|\lambda_s)p(\lambda_s)}{p(X)} \nonumber \\
			& \propto & \argmax_{s \in [1,S]}p(X|\lambda_s) \nonumber \\
			& \propto & \argmax_{s \in [1,S]}\sum^{K}_{k=1}\mathrm{log}p(\mathbf{x}_{k}|\lambda_s).
\end{eqnarray}
The first equation is due to Bayes' rule. The first proportion is assuming $p(\lambda_s) = 1/S$ and $p(X)$ is the same for all accent models. The second proportion uses logarithm and independence between input samples $\mathbf{x}_k$, $k \in [1,K]$.

\subsection{LDA and HLDA} 
\label{subsec:hlda}

Compared with Principle Component Analysis (PCA), which transforms data into eigenspace and preserves the data dimensions with larger variation \cite{ge2011pca}, Linear Discriminant Analysis (LDA) reduces dimensions by mapping data into a subspace while maximizing the discriminative information. Assume there are $K = \sum_{s=1}^{S}K_{s}$ number of $M$-dimensional data vectors $\bm{x}_k$ in $S$ classes, where $K_{s}$ is the number of vectors in class $s \in [1,S]$. Let the global mean $\bm{\Phi}$ over all classes be $\bm{\Phi} = \frac{1}{K}\sum_{k}^{K}\bm{x}_k$ and the local mean $\bm{\Phi}_s$ for each class $s$ be $\bm{\Phi_s} = \frac{1}{K_s}\sum_{\bm{x}_{k} \in s}\bm{x}_{k}$
%
%
respectively. Then, we define between-class scatter $S_B$ and within-class scatter $S_W$ by 
%
%
\begin{eqnarray} \label{eq:bcs}
	S_B &=& \frac{1}{K}\sum^{K}_{k=1}(\bm{x}_k-\bm{\Phi})(\bm{x}_k-\bm{\Phi})^{T} ,
\end{eqnarray}
%
\begin{eqnarray} \label{eq:wcs}
	S_W &=& \frac{1}{S}\sum^S_{s=1}\sum_{\bm{x}_k \in s}(\bm{x}_k-\bm{\Phi}_s)(\bm{x}_k-\bm{\Phi}_s)^{T} .
\end{eqnarray}
%

If we choose $\mathbf{w}$ from the underlying space $W$, then $\mathbf{w}^T S_B \mathbf{w}$ and $\mathbf{w}^T S_W \mathbf{w}$ are the projections of $S_B$ and $S_W$ onto the direction $\mathbf{w}$. Searching the directions $\mathbf{w}$ for the best class discrimination is equivalent to maximizing the ratio of $(\mathbf{w}^T S_B \mathbf{w}) / ({\mathbf{w}^T S_W \mathbf{w}})$ subject to $\mathbf{w}^T S_W \mathbf{w} = 1$. The latter is called the Fisher Discriminant Function and can be converted to 
%
%
by Lagrange multipliers and solved by eigen-decomposition of $S^{-1}_W S_B$. By selecting eigenvectors associated with the most significant $m$ eigenvalues of $S^{-1}_W S_B$, one can map the original $M$-dimensional data into a $m$-dimensional subspace for discriminative feature reduction.  

\begin{figure}[tb]
  \centering
  \includegraphics[scale=0.3]{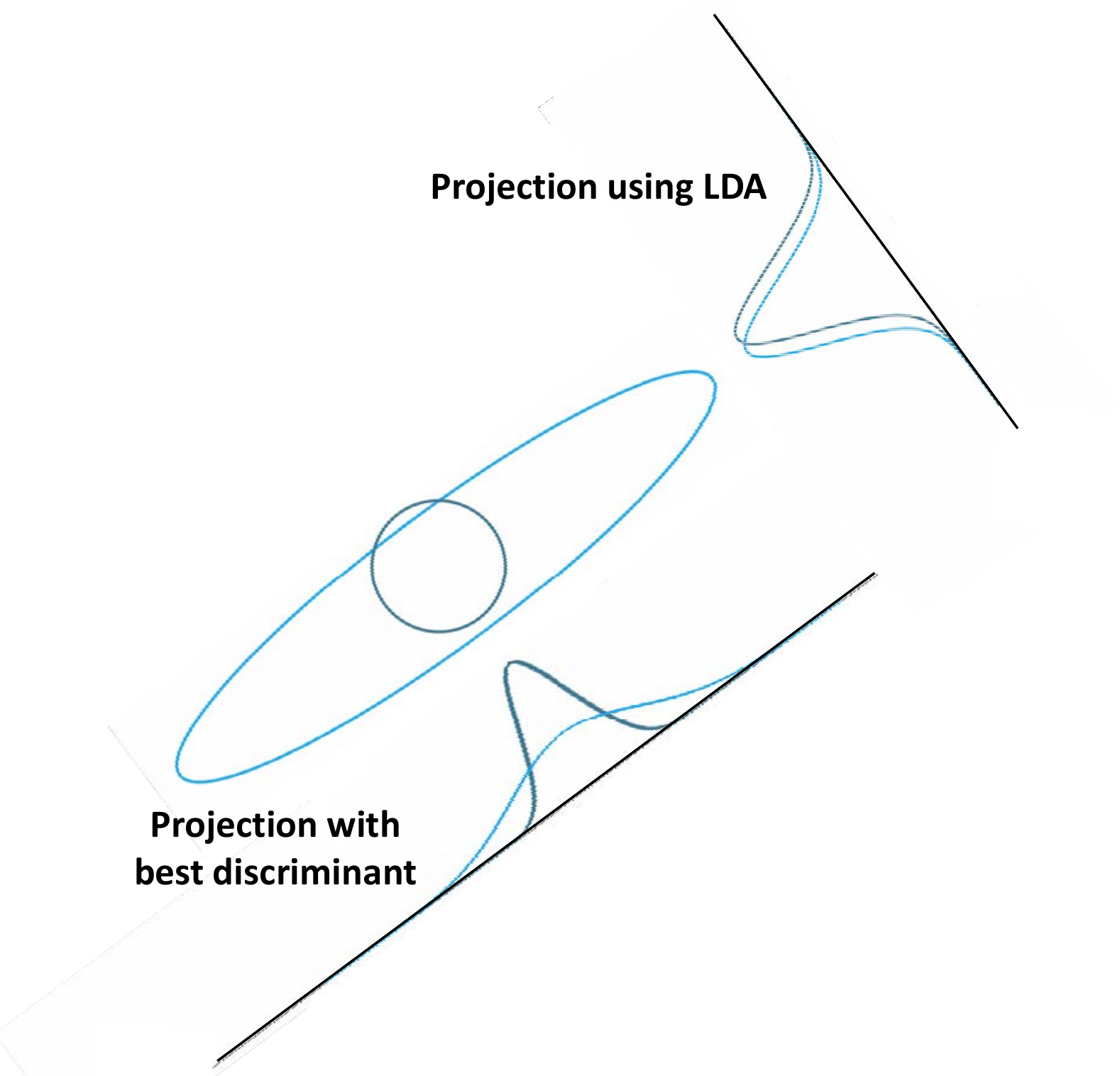}
  \caption{How LDA fails with two Gaussian distributions}
  \label{fig:demo_hlda}
  \myfigureshrinker
\end{figure}

LDA is derived with the assumption that features in various dimensions have the same variance, which may not be the case in the real problem. For example, consider two classes of data with the Gaussian distributions shown in Fig. \ref{fig:demo_hlda}. They have the  same variance and slightly different means in one direction, while same mean and  significantly different variances in the other distribution. LDA will project the data to the first direction, since it maximizes the ratio of between-class scatter $S_B$ and within-class scatter $S_W$. However, the other direction will lead to the best discriminant information in this case.
This work uses Kumar's method \cite{Kumar:1997investigation} to generalize LDA to HLDA using Maximum Likelihood Estimation (MLE) on Gaussian distributions.


\subsection{Results for Baseline}
\label{subsec:baseline}

\begin{figure}[tb]
  \centering
  \includegraphics[scale=0.35]{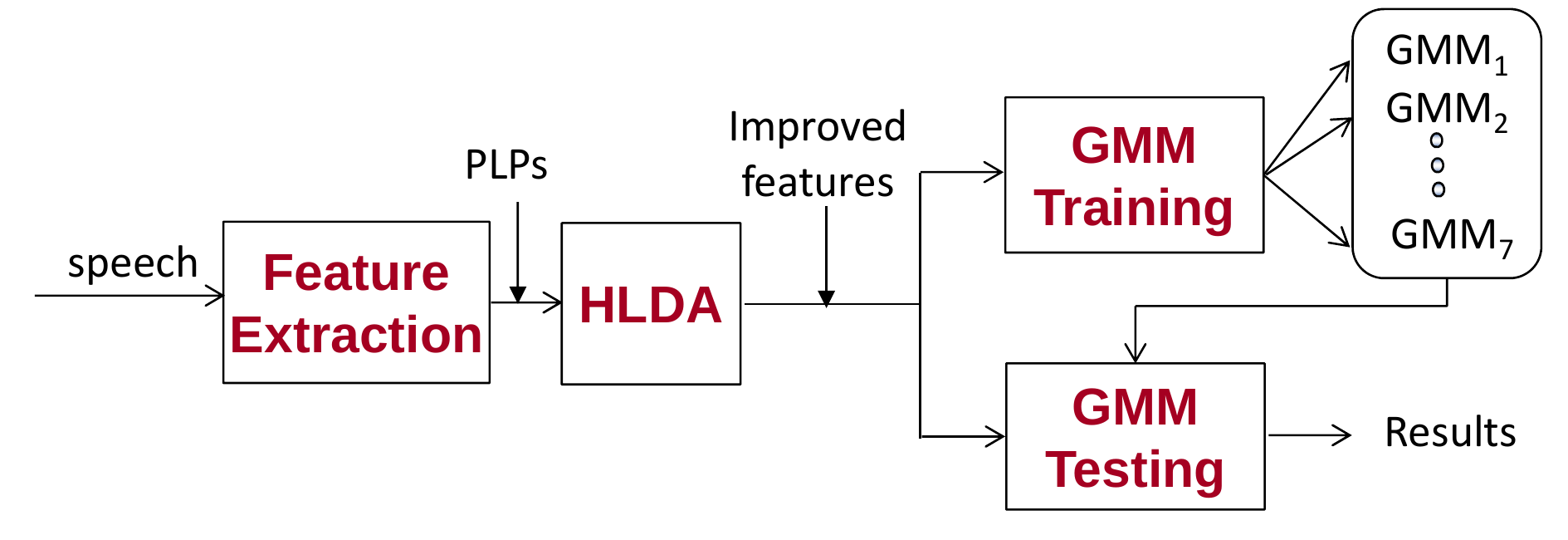} 
  \caption{Diagram of accent classification based on pure acoustic information}
  \label{fig:accentdetect_base}
  \myfigureshrinker
\end{figure}

The diagram of the 7-way accent classification based on pure acoustic information is demonstrated in Fig. \ref{fig:accentdetect_base}.
PLP-HLDA features with context-size 1 and reduced dimension 20 is used. The context-size factor is used to duplicate features for potential performance improvement. For example, with context-size 1, the original feature frame is elongated with the concatenation from its 1 left frame and 1 right frame. Both the GMM classifier and the improved GMM-HLDA classifier were trained with features of various types of accents of 256 Gaussian mixtures. These parameters, including order of GMM, feature dimension in PLP and HLDA, and context-size were optimized with development set. The performance on the testing set achieve 40\% and 46\% accuracies using GMM classifier and GMM-HLDA classifier. 

\vspace{-0.1cm}
\section{Improved with Vowel Representation}
\label{sec:improved}

To construct the classifier with vowel representation, instead of directly measuring vowel shifting from standard speech to accented one, the same vowel of various types of accents are trained as separated GMMs; instead of using only the fundamental 5 vowels described in Sec.  \ref{sec:vowel_representation}, the same concept is generalized and all 15 vowels in Arpabet \cite{arpabet} listed in Tab. \ref{tab:vowels} are used.  

\begin{table}[tb]
\centering
\footnotesize
\caption{Vowels in Arpabet}
\label{tab:vowels}
\renewcommand{\tabcolsep}{3pt}
\renewcommand\baselinestretch{0.5}\selectfont
\begin{tabular}{@{} {l}*{9}{l} @{}} \toprule
\textbf{Vowel} & aa & ae & ah & ao & aw & ay & eh & er \\ 
\textbf{Example} & f\textbf{a}ther & f\textbf{a}st & s\textbf{u}n & h\textbf{o}t & h\textbf{ow} & m\textbf{y} & r\textbf{e}d & b\textbf{ir}d  \\\midrule
\textbf{Vowel} & ey & ih & iy & ow & oy & uh & uw & \\
\textbf{Example} & s\textbf{ay} & b\textbf{i}g & m\textbf{ee}t & sh\textbf{ow} & b\textbf{oy} & b\textbf{oo}k & f\textbf{oo}d \\\bottomrule
\end{tabular}
\myfigureshrinker
\end{table}

Given $S$ types of accents and $T$ numbers of vowels, $X^{(t)}$ is the extracted feature set for $t^\textrm{th}$ vowel, the improved GMM classifer as the combination of GMM classifers of all vowels, can be formulated as:
\begin{eqnarray} \label{eq:vrclassifier}
	\hat{S} & = & \argmax_{s \in [1,S]}\sum_{t=1}^{T}w_{t}p(\lambda_{s,t}|X^{(t)}) \nonumber \\
			& \propto & \argmax_{s \in [1,S]} \sum_{t=1}^{T}w_{t}\sum^{K}_{k=1}\mathrm{log}p(\mathbf{x}_{k}^{(t)}|\lambda_{s,t}).
\end{eqnarray}
where $\lambda_{s,t}$ is the GMM for $s^\textrm{th}$ accent and $t^\textrm{th}$ type of vowels, and $w_{t}$ is the proportion of $t^\textrm{th}$ vowel in the whole vowel set.

Adding this additional layer on the GMM classifier is critical for finding the vowel sets which preserve the accents and is shown to improve on classifying accents. However, it requires recognizing these vowels in the front end. During training and development, the phoneme alignment is performed to extract vowels, while during testing, a subset of the recognized vowels with certain level of confidence are selected after phoneme recognition. The HTK Speech Recognition Toolkit was used here for the phoneme alignment and recognition with triphone acoustic models.

\subsection{Phoneme Alignment and Recognition}

\begin{figure*}[ht!]
  \centering
  \includegraphics[height=3.45cm, width=11cm]{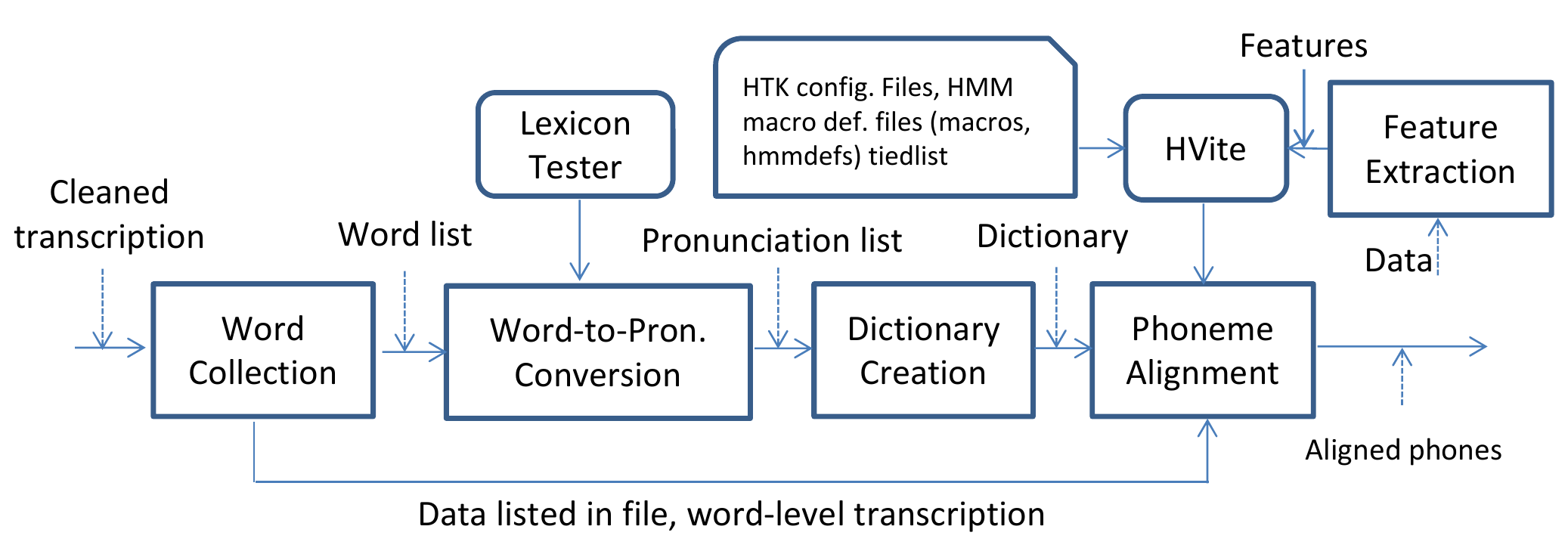}               
  \caption{Dictionary preparation and phoneme alignment for FAE corpus}
  \label{fig:dppa}
  \myfigureshrinker
\end{figure*}

In the system developement, with the partial in-house transcriptions of the speech from 7 major accents, we prepare dictionary needed for phoneme alignment using HVite in HTK. Fig. \ref{fig:dppa} demonstrates the process of dictionary preparation and phoneme alignment for  FAE. 
The dictionary file is a list of word-pronunciations pairs in HTK format, which can be obtained through the process of word collection, word-to-pronunciation conversion with an in-house lexicon tester and HTK dictionary file creation. In the phoneme alignment, the HTK configuration file, the HMM model definition and the tired list are all trained using Fisher corpus.

In the system test, since there is no transcription available , in order to find features corresponding to vowels, accented speech is recognized using HTK and only a subset of recognized vowels with certain level of confidence based on the $n$-gram log likelihood are used. This threshold is predefined with training and development data.   

\subsection{Results for Improved Classifier}
\label{subsec:improved}

Here 39-dimensional PLP features with MVNs are used in the implementation of accent classification. After training GMMs on seperated vowels, GMMs of 7 vowels out of 15 of each accent are selected to form the mixed GMM classifer for that accent. The overall classification accuracy is 50.9\%, which gains 4.5\% improvement from the GMM classifer trained with HLDA features. Tab. \ref{tab:result_accentclassification} compares the performances of all three methods. The combination of classifier and features include a) GMMs with PLP, trained per accent; b) GMMs with HDLA (20 dimensions with context size 1, optimized from 39-dimensional PLP), trained per accent; and c) GMMs with HLDA, trained per accent and per vowel.  
The accuracy is obtained from accented speech about 20-second duration, and is competitive compared with the state-of-art results in \cite{choueiter2008empirical}, \cite{omar2010novel} and \cite{macias2003acoustic}.
\begin{table}[htb]
\centering
\footnotesize
\caption{Result comparison of 7-way accent classification}
\label{tab:result_accentclassification}
\renewcommand{\tabcolsep}{2pt}
\renewcommand\baselinestretch{1}\selectfont
\begin{tabular}{@{} llll @{}} \toprule
Model & $\mathbf{GMM}_{base}^{256}$ & $\mathbf{GMM}_{base}^{256}$ & $\mathbf{GMM}_{vowel}^{256}$ \\
Feature & $\mathbf{PLP}_{MVN}^{39}$ & $\mathbf{HLDA}_{C1}^{20}$ & $\mathbf{HLDA}_{C1}^{20}$ \\
Accuracy & 40.3\% & 46.4\% & 50.9\% \\\bottomrule
\end{tabular}
\myfigureshrinker
\end{table}

\section{Summary and Future Work}
\label{sec:conclusion}

This work shows the classification accuracy improvement with HLDA feature optimization and extracted vowels from accented speech for developing GMM classifiers. There are at least several areas that can be addressed for further improvement. First, more sophisticated classifiers such as deep neural network classifier \cite{graves2013speech, ge2015sleep} may also be used for accent classification. Second, since the data for each accent is very limited, a universal classifier based on Restricted Boltzmann Machine (RBM), instead of traditional GMMs for each accents can be explore \cite{larochelle2008classification}. RBM is trained using data of all accents, with capability to deviate with different accents. Third, accent clustering based on certain distance measurements, such as Bhattacharyya distance \cite{Bhattacharyya:1946} can also be used to pre-classify accents into several clusters, which may potentially help narrow down the search scope and improve the classification accuracy. Forth, in the triphone phoneme alignment and recognition, currently all triphones with the same mid-phone are treated the same. However, the accent patterns may stay in the transition of phonemes, which can be investigated later. 

\vspace{-0.1cm}
{\small
\bibliographystyle{ieee}
\bibliography{paper_acpr3}
}

\end{document}